\documentstyle[]{article}

\newfam\msbfam
\font\tenmsb=msbm10                     \textfont\msbfam=\tenmsb
\font\sevenmsb=msbm7            \scriptfont\msbfam=\sevenmsb
\font\fivemsb=msbm5                     \scriptscriptfont\msbfam=\fivemsb
\def\Bbb{\fam\msbfam \tenmsb}

\newtheorem{th}{Theorem}
\newtheorem{ax}{Axiom}
\newtheorem{lm}{Lemma}
\newtheorem{df}{Definition}
\newtheorem{pr}{Proposition}
\newtheorem{cl}{Corollary}
\newtheorem{as}{Assumption}
\newtheorem{qu}{Question}
\newcommand{\bth}{\begin{th}\hspace{-5pt}{\bf .} \ }
\newcommand{\eth}{\end{th}}
\newcommand{\bax}{\begin{ax}\hspace{-5pt}{\bf .} \ }
\newcommand{\eax}{\end{ax}}
\newcommand{\blm}{\begin{lm}\hspace{-5pt}{\bf .} \ }
\newcommand{\elm}{\end{lm}}
\newcommand{\bdf}{\begin{df}\hspace{-5pt}{\bf .} \ }
\newcommand{\edf}{\end{df}}
\newcommand{\bpr}{\begin{pr}\hspace{-5pt}{\bf .} \ }
\newcommand{\epr}{\end{pr}}
\newcommand{\bcl}{\begin{cl}\hspace{-5pt}{\bf .} \ }
\newcommand{\ecl}{\end{cl}}
\newcommand{\bas}{\begin{as}\hspace{-5pt}{\bf .} \ }
\newcommand{\eas}{\end{as}}
\newcommand{\bqu}{\begin{qu}\hspace{-5pt}{\bf .} \ }
\newcommand{\equ}{\end{qu}}

\newcommand{\bit}{\begin{itemize}}
\newcommand{\eit}{\end{itemize}\par\noindent}
\newcommand{\beq}{\begin{equation}}
\newcommand{\eeq}{\end{equation}\par\noindent}
\newcommand{\beqa}{\begin{eqnarray*}}
\newcommand{\eeqa}{\end{eqnarray*}\par\noindent}
\newcommand{\beqn}{\begin{eqnarray}}
\newcommand{\eeqn}{\end{eqnarray}\par\noindent}

\newcommand{\bla}{\left\{\begin{array}{l}}
\newcommand{\ela}{\end{array}\right.}
\newcommand{\bra}{\left\{\begin{array}{r}}
\newcommand{\era}{\end{array}\right.}

\newcommand{\ben}{\begin{enumerate}}
\newcommand{\een}{\end{enumerate}\par\noindent}

\begin{document}


\noindent\centerline{\large{\bf FORCING DISCRETIZATION AND DETERMINATION}} 
 
\smallskip\noindent\centerline{\large{\bf IN QUANTUM HISTORY
THEORIES}}

\bigskip\par\noindent
\medskip\par\noindent
\centerline{\normalsize{Bob Coecke}} 
\par\medskip\noindent
\centerline{\footnotesize{Imperial College of Science, Technology \&
Medicine, Theoretical Physics Group,}}\vspace{-1mm} 
\par\noindent 
\centerline{\footnotesize{The Blackett Laboratory, South  Kensington, LondonSW7
2BZ;}}
\par\smallskip\noindent 
\centerline{\footnotesize{Free University of
Brussels (VUB),  Department of Mathematics (FUND),}}\vspace{-1mm}
\par\noindent 
\centerline{\footnotesize{Pleinlaan 2,
B-1050 Brussels, Belgium (permanent mailing address)\,;}}\vspace{-1mm} 
\par\noindent 
\centerline{\footnotesize{e-mail: bocoecke@vub.ac.be\,.}} 
\par\bigskip\noindent
\centerline{Contribution to the conference:}
\centerline{Foundations of Probability and Physics, Vaxjo, Sweden, November 2000\,;}
\centerline{To appear in {\it Quantum  Probability and Related Topics}
{\bf IX},}
\centerline{L. Accardi, Ed., World Scientific.}

\begin{abstract}
We present a formally deterministic 
representation for quantum history theories
where we obtain the probabilistic structure via a discrete contextual
variable: no continuous probabilities are as such involved at the 
primal level --- we conceive as a history theory any theory that deals 
with sequential quantum measurements but
remains essentially a dichotomic propositional theory.  A major part of the paper consists
of a concise survey of arXiv: quant-ph/0008061 and quant-ph/0008062{\,}.
\end{abstract}  

\section{{\large INTRODUCTION}} 

In this paper we propose and study a model for history theories in 
which the probability
structure emerges from a finite number of contextual happenings, any 
next happening having
a fixed chance to occur under the condition that the previous one happened.
Although this model cannot have a canonical mathematical status since 
it has been proved that
this type of representation in general admits no essentially unique 
``smallest one''
\cite{C97}$^{ii}$, it provides insight in the emergence of logicality 
in the ``History Projection
Operator'' setting \cite{I94}, and it illustrates how deterministic 
behavior can be encoded beyond
those interpretations of quantum history theories that are 
interpretationally restricted by
so-called consistency or quasi-consistency (e.g., approximate 
decoherence).  The
particular motivation for this ``paradigm case study'' finds its 
origin in structural
considerations towards a theory of quantum gravity 
\cite{A95,I97,R98}. As argued in
\cite{IB00}, although the relative frequency interpretation of 
probability justifies the
continuous interval as the codomain for value assignment, ``... in 
the quantum gravity
regime standard ideas of space and time might break down in such a 
way that the idea of
spatial or temporal `ensembles' is inappropriate. For the other main 
interpretations of
probability --- subjective, logical, or propensity --- there seems to 
be no compelling {\it
a priori} reason why probabilities should be real numbers.'' Our 
model should be envisioned
as a deconstructive step unraveling the probabilistic continuum as it 
appears in standard
quantum theory, reducing it explicitly to a discrete temporal 
sequence of (contextual)
events.  The as such emerging temporal sequence is then easier to 
manipulate towards
alternative encoding of contextual events, e.g., in propositional terms.
It also enables a separate treatment of internal (the system's) and 
external (the
context's) time-encoding variable.

Although quantum history theories
are currently most frequently envisioned in a context of so-called 
decoherence we prefer to take
the minimal perspective that a history theory is a theory that deals 
with sequential quantum
measurements but
remains essentially a dichotomic propositional theory.  This is 
formally encoded in a
rigid way in the History Projection Operator-approach \cite{I94}. We 
also mention
recently studied sequential structures in the context of quantum 
logic, of which references can
be found in \cite{CMS01}, resulting in a dynamic disjunctive quantum 
logic, which provides an
appropriate formal context to discuss the logicality of history theories.

A general theory on deterministic contextual models can be found in 
\cite{C97}. Note here that
what we consider as contextuality is that in a measurement there is 
an interaction between the
system and its context and that precisely this interaction to some 
extend may influence the
outcome of a measurement.  A lack of knowledge on the precise 
interaction then yields
quantum-type uncertainties
\cite{A86}. Besides this interpretational issue, classical 
representations are important since
we think classical, so even without giving any
conceptual significance to the representation, it provides a mode to 
think deterministically
in terms of determined trajectories of the system's state, without having to
reconcile with concrete non-canonical constructs like pilot-wave mechanics.

\section{{\large OUTCOME DETERMINATION VIA CONTEXTUAL\\ MODELS}}

We will present the required results in full abstraction such that 
the reader clearly sees which
structural ingredient of quantum theory determines existence of 
contextual models. For
details and proofs we refer to \cite{C97}.  Let ${\cal B}({\Bbb 
R}^\nu)$ denote the
Borel subsets of ${\Bbb R}^\nu$\,.
\bdf
A `probabilistic measurement system' is given by\,:
\par\noindent
(i) A set of states $\Sigma$ and a set of measurements ${\cal E}$\,;
\par\noindent
(ii) For each $e\in{\cal E}$ an outcome set $O_e\in{\cal B}({\Bbb R}^\nu)$\,, a
$\sigma$-field
${\cal B}(O_e)$ of $O_e$-subsets and (Kolmogorovian) probability 
measures $P_{p,e}:{\cal
B}(O_e)\to[0,1]$ for each $p\in\Sigma$\,.
\edf
The canonical example is that of quantum theory with every Hilbert 
space ray $\psi$
representing a state, every self-adjoint operator $H$ representing a 
measurement with its
spectrum
$O_H\subseteq{\Bbb R}$ as outcome set where the $\sigma$-structure   
${\cal B}(O_H)$ is
inherited from that of
${\cal B}({\Bbb R})$ and with probability measures
$P_{\psi,H}(E):=\langle\psi|P_E\psi\rangle$ where
$P_E$ denotes the spectral projector for $E\in{\cal B}(O_H)$\,.
In benefit of insight and also for notational convenience we will 
from now on assume
that the measurements $e\in{\cal E}$ are represented in a one to one 
way by their outcome sets
$O_e$ --- note that whenever ${\cal E}$ can be represented by points 
of ${\Bbb R}^{\nu'}$ it
then suffices to consider
${\Bbb R}^\nu\times{\Bbb R}^{\nu'}={\Bbb R}^{\nu+\nu'}$ in stead of 
${\Bbb R}^{\nu}$ to fulfill
this assumption, taking
$O_e\times\{e\}$ as the corresponding outcome set.  We stress however 
that the results listed
below also hold in absence of this assumption \cite{C97}$^{i,ii}$.
\bdf
A `pre-probabilistic hidden measurement system' is given by\,:
\par\noindent
(i) A set of states $\Sigma$ and a set of measurements ${\cal E}$\,;
\par\noindent
(ii) Sets ${\cal O}\subseteq{\cal B}({\Bbb R}^\nu)$ and $\Lambda$ 
that parameterize
${\cal E}$\,, i.e., ${\cal E}=\{e_{\lambda,O}|\lambda\in\Lambda, 
O\in{\cal O}\}$\,, and
each $e\in{\cal E}$ goes equipped with a map $\varphi_{\lambda, 
O}:\Sigma\to O$\,.
\edf
We can represent $\{\varphi_{\lambda, O}|\lambda\in\Lambda\}$ as
$\varphi_O:\Sigma\times\Lambda\to 
O:(p,\lambda)\mapsto\varphi_{\lambda,O}(p)$\, giving
$\Lambda$ a similar formal status as the set of states $\Sigma$\,, or as
$\Delta\Lambda_O:\Sigma\times{\cal B}(O)\to{\cal
P}(\Sigma):(p,E)\mapsto\{\lambda|\varphi_O(p,\lambda)\in E\}$
where ${\cal P}(\Lambda)$ denotes the set of subsets of $\Lambda$\,.
{\it The core of this definition is that given a state $p\in\Sigma$ and a value
$\lambda\in\Lambda$ we have a completely determined outcome 
$\varphi_O(p,\lambda)$\,.
These pre-probabilistic hidden measurement systems encode as such 
fully deterministic settings.}
\bdf
Whenever for a given pre-probabilistic hidden measurement system 
$\bigl(\Sigma, {\cal
E}({{\cal O},\Lambda}), \{\varphi_O\}_{O\in{\cal O}}\bigr)$ there 
exists a $\sigma$-field
${\cal B}(\Lambda)$ of $\Lambda$-subsets that satisfies
$\bigcup_{O\in{\cal O}}\{\Delta\Lambda_O(p,E)|(p,E)\in\Sigma\times{\cal
B}(O)\}\subseteq{\cal B}(\Lambda)$, it defines a `probabilistic 
hidden measurement system' if
a probability measure
$\mu:{\cal B}(\Lambda)\to[0,1]$ is also specified.
\edf
The condition on $\Delta\Lambda$ requires that all 
$\Delta\Lambda_O(p,E)$ are ${\cal
B}(\Lambda)$-measurable, such that to all triples $(p,O,E)$ we can 
assign a value
$P_{p,O}(E):=\mu\bigl(\Delta\Lambda_O(p,E)\bigr)\in[0,1]\,.$
As such, any probabilistic hidden measurement system defines a 
measurement system.
The question then rises whether every probabilistic measurement 
system (MS) can be encoded as
a  probabilistic hidden measurement system (HMS).  The answer to this 
question is yes \cite{C97}$^{i}$,
\S4.2, Theorem 1,2 \& 3\,: {\it There always exists a canonical 
HMS-representation for
$\Lambda\cong[0,1]$\,,
${\cal B}(\Lambda)\cong{\cal B}([0,1])$ (i.e., the Borel sets in 
$[0,1]$\,) and{\,}
$\mu_u([0,a]):=a$\,, i.e., uniformly distributed\,} --- the proof goes via a
construction using the Loomis-Sikorski Theorem \cite{L47,S48} and 
Marczewski's Lemma
\cite{HT48}. It makes as such sense to  investigate how the different
possible HMS-representations for different non-isomorphic pairs
$\bigl({\cal B}(\Lambda),\mu\bigr)$ are structured --- below it will 
become clear what we mean
here by non-isomorphic. First we will discuss an example that 
illustrates the above\,; it
traces back to \cite{A86} and details and illustrations can be found in
\cite{A93,C97}$^{ii}$\,.  Consider the states of a spin-${1\over 2}$ 
entity encoded as a point on the
Poincar\'e sphere $\Sigma_\circ(\cong{\Bbb C}^2/{\Bbb 
C})\subseteq{\Bbb R}^3$\,. Then any pair
of antipodically located points of $\Sigma_\circ$ encodes mutual 
orthogonal states, as such
encodes mutual orthogonal one-dimensional projectors and thus a 
(dichotomic) measurement.
Let $p\in\Sigma_\circ$\,, let $(\alpha,
\neg\alpha)$ be a pair of mutual orthogonal points of $\Sigma_\circ$ 
and let $\Lambda$ be the
diagonal connecting
$\alpha$ and $\neg\alpha$\,. Let $x_p\in\Lambda$ be the orthogonal 
projection of $p$ on
the diagonal $\Lambda$\,. Then, for $\lambda\in[x_p,\neg\alpha]$\,, i.e.,
$x_p\in[\alpha,\lambda]$\,, we set
$\varphi(p,\lambda)=\alpha$ and for $\lambda\in[\alpha,x_p[$\,, i.e.,
$x_p\in]\lambda,\neg\alpha]$\,, we set
$\varphi(p,\lambda)=\neg\alpha$\,. One then verifies that for $\mu_\circ:={\cal
B}([\alpha,\neg\alpha])\to[0,1]:[\alpha,(1-x)\alpha+x\neg\alpha]\mapsto 
x$\,, i.e., uniformly
distributed, we obtain exactly the probability structure for 
spin-${1\over2}$ in quantum
theory\,.\footnote{As shown in \cite{C95a,C98} this deterministic 
model for spin-${1\over 2}$ in
${\scriptstyle{\Bbb R}}^3$ can be generalized to ${\scriptstyle{\Bbb 
R}}^3$-models for
arbitrary spin-${N/2}$\,. The states are then represented in the so 
called Majorana
representation \cite{M32,B74}, i.e., as
$N$ copies of $\Sigma_\circ$\,. Correct probabilistic behavior is 
then obtained by introducing
entanglement between the $N$ different ``spin-${1\over 2}$ 
systems''.} An interpretational
proposal of this model could be the following:\cite{A86,A93,A94} {\it 
Rather than decomposing
states as in so-called hidden variable theories, here we decompose 
the measurements in
deterministic ones --- the probability measure $\mu$ should then be 
envisioned as encoding
the lack of knowledge on the interaction of the measured system with 
its environment,
including measurement device.}

We now introduce a notion of ``relative size'' of 
HMS-representations, justifying the
use of ``smaller''.  Given a $\sigma$-algebra\,\footnote{I.e., a 
``pointless'' $\sigma$-field.
In particular, it follows from the Loomis-Sikorski theorem 
\cite{L47,S48} that all
separable
$\sigma$-algebras (i.e., which contain a countable dense subset) can 
be represented as a
$\sigma$-field  --- it as such also follows that assuming that ${\cal 
B}(\Lambda)$ is a
$\sigma$-field and not an abstracted $\sigma$-algebra imposes no 
formal restriction.} and
probability measure
$\mu:{\cal B}\to[0,1]$ denote by ${\cal B}/\mu$ the $\sigma$-algebra 
of equivalence classes
$[E]$ with respect to the relation $E\sim
E'\Leftrightarrow\mu(E'\cap E^c)=\mu(E\cap E'^c)=0$\,, i.e., iff $E$ 
and $E'$ coincide up to a
symmetric difference of measure zero\,. The ordering of ${\cal 
B}/\mu$ is inherited from
${\cal B}$\,. For notational convenience denote the induced measure ${\cal
B}/\mu\to[0,1]:[E]\mapsto\mu(E)$ again by
$\mu$\,.  Given two  pairs $({\cal B},\mu)$ and $({\cal B}',\mu')$ 
consisting of
separable $\sigma$-algebras and probability measures on them set\,:

\smallskip
$\bullet$ $
({\cal B},\mu)\leq({\cal B}',\mu')\Leftrightarrow
\bla
\exists f:{\cal B}/\mu\to{\cal B}'/\mu'\,,\ {\rm an}\ {\rm 
injective}\ \sigma{\rm
\mbox{-}morphism}\\
\mu'\circ f=\mu
\ela
$

\smallskip\noindent
We call $({\cal B},\mu)$ and $({\cal B}',\mu')$ equivalent, denoted 
$({\cal B},\mu)\sim({\cal
B}',\mu')$\,, whenever in the above $f$ is a
$\sigma$-isomorphism. Given two MS $(\Sigma, {\cal E})$ and 
$(\Sigma', {\cal E}')$ we set\,:

\smallskip
$\bullet$ $
(\Sigma, {\cal E})\,\sim_{_{MS}}\!\!(\Sigma', {\cal E}')\Leftrightarrow
\bla
\exists s:\Sigma\to\Sigma'\,, \exists t:{\cal E}\to{\cal E}'\,,
{\rm both}\ {\rm bijections}\\
\forall e\in{\cal E},\exists f_e:{\cal B}(O_e)\to{\cal 
B}(O_{t(e)})\,, {\rm a}\ \sigma{\rm
\mbox{-}isomorphism}\\
\forall p\in\Sigma,\forall e\in{\cal E}:P_{s(p),t(e)}\circ f_e=P_{p,e}
\ela
$

\smallskip\noindent
Via this equivalence relation we can define a relation $\leq_{_{MS}}$ 
between classes of
measurement systems ${\cal M}$ and ${\cal M}'$ as ${\cal 
M}\,\leq_{_{MS}}\!\!{\cal M}'$ if
for all $(\Sigma, {\cal E})\in{\cal M}$ there exists $(\Sigma', {\cal 
E}')\in{\cal M}'$ such
that
$(\Sigma, {\cal E})\,\sim_{_{MS}}\!\!(\Sigma', {\cal E}')$\,, i.e., 
if ${\cal M}$ is included
in ${\cal M}'$ up to MS-equivalence. We can then prove the following\,:

\smallskip\noindent
(i) $({\cal B},\mu)\sim({\cal B}',\mu')$ if and only if $({\cal 
B},\mu)\leq({\cal B}',\mu')$
and $({\cal B}',\mu')\leq({\cal B},\mu)$ --- \cite{C97}$^{ii}$, \S3, 
Lemma 1\,; thus, the equivalence
classes with respect to
$\sim$ constitute a partially ordered set (poset) for the ordering 
induced by $\leq$\,; we
will denote the set of these equivalence classes by ${\bf M}$\,, a 
class in it will be denoted
via a member of it as $[{\cal B},\mu]$\,.

\smallskip\noindent
(ii) When setting ${\bf M}_{_{HMS}}:=\bigl\{{\cal M}[{\cal 
B}(\Lambda),\mu]\bigm|[{\cal
B}(\Lambda),\mu]\in{\bf M}\bigr\}$ where ${\cal M}[{\cal 
B}(\Lambda),\mu]$ stands for all HMS
with ${\cal B}(\Lambda')$ and $\mu'$ such that $({\cal
B}'(\Lambda'),\mu')\in[{\cal B}(\Lambda),\mu]$\,, we have that
$({\cal
B}(\Lambda),\mu)\leq({\cal B}'(\Lambda'),\mu')$ and ${\cal M}[{\cal
B}(\Lambda),\mu]\leq_{_{MS}}\!{\cal M}[{\cal B}'(\Lambda'),\mu']$ are 
equivalent \cite{C97}$^{ii}$, \S3,
Theorem 2\,.
This then results in\,:
\bth\label{th:poset}
$({\bf M},\leq)$ and $({\bf M}_{_{HMS}},\leq_{_{MS}}\!)$ are isomorphic posets.
\eth
One of the crucial ingredients in (ii) above and also in the proof 
for general existence with
$\Lambda\cong[0,1]$ is the following: when setting
$\Delta{\bf M}(\Sigma,{\cal E}):=\{({\cal 
B}(O_e),P_{p,e})|p\in\Sigma, e\in{\cal E}\}$\,, we
obtain that $\Sigma,{\cal E}$ admits a HMS-representation with
${\cal B}(\Lambda)$ and $\mu$ if and only if
$\Delta{\bf M}(\Sigma,{\cal E})\leq({\cal B}(\Lambda),\mu)$\,, where 
the order applies
pointwisely to the elements of $\Delta{\bf M}(\Sigma,{\cal E})$ 
\cite{C97}$^{i}$, \S4.2, Theorem 1\,.
Using this and Theorem \ref{th:poset} above we can now translate
properties of ${\bf M}$ to propositions on the existence of certain 
HMS-representations.
We obtain the following\,:
\par\smallskip\noindent
(i) $({\bf M},\leq)$ is not a join-semilattice, thus\,: {\it In 
general there exists no
smallest HMS-representation.}  As such we will have to refine our 
study to particular
settings where we are able to make statements whether there exists a 
smallest one,
and if not, whether we can say at least something on the cardinality 
of $\Lambda$.
\par\smallskip\noindent
(ii) One can prove a number of criteria on $\Delta{\bf 
M}(\Sigma,{\cal E})$ that force
$({\cal B}(\Lambda),\mu)\sim\bigl({\cal B}([0,1]),\mu_u\bigr)$\, as 
such assuring
existence of a smallest representation. Among these the following.
Let ${\bf M}_{finite}:=\bigl\{\bigl({\cal B}(X),\mu\bigr)\in{\bf 
M}\bigm|X\ {\rm is}\
{\rm finite}\bigr\}$\,. {\it If ${\bf M}_{finite}\subseteq\Delta{\bf 
M}(\Sigma,{\cal E})$
than $\Lambda$ cannot be discrete.} It then follows for example that 
quantum theory restricted
to measurements with a finite number of outcomes still requires 
$\Lambda\cong[0,1]$\,.
\par\smallskip\noindent
(iii) Let ${\bf M}_N:=\bigl\{\bigl({\cal B}(X),\mu\bigr)\in{\bf 
M}\bigm|X\ {\rm has}\ {\rm
at\ most}\ N {\rm\ elements}\bigr\}$\,. {\it If $\Delta{\bf 
M}(\Sigma,{\cal E})\subseteq{\bf
M}_N$ then there exists a HMS-representation
with $\Lambda\cong{\Bbb N}$\,.} Thus, quantum theory restricted to 
those measurements with at
most a fixed number $N$ of outcomes has discrete HMS-representation.
\par\smallskip\noindent
(iv) {\it If $\Delta{\bf M}(\Sigma,{\cal E})={\bf M}_N$ then there 
exists no smallest
HMS-representation. Neither does it exist when fixing the number of outcomes.}
So there is no essentially unique smallest HMS-representation for 
$N$-outcome quantum theory.
\par\smallskip\noindent
Although there exists no smallest and as such no canonical discrete 
HMS-representation we will
give the construction of one solution for dichotomic (or 
propositional) quantum theory, i.e.,
$N=2$\,, since this will constitute the core of the model presented 
in this paper. We
will follow \cite{C97}$^{iii}$, to which we also refer for a 
construction for arbitrary $N$\,.
Let us denote the quantum mechanical probability to obtain a positive 
outcome in a
measurement of a proposition or question $\alpha$ on a system in 
state $p$ as $P_p(\alpha)$ ---
the outcome set consists here of ``we obtain a positive answer for 
the question $\alpha$'',
slightly abusively denoted as
$\alpha$ itself, and ``we obtain a negative answer for the question 
$\alpha$'', denoted as
$\neg\alpha$\,.  Set inductively for $\lambda\in{\Bbb 
N}$\,:\,\footnote{We agree on ${\Bbb
N}:=\{1,2,\ldots\}$\,. Note here that already by non-uniqueness of binary
decomposition --- ${1\over2}={1\over 2^1}=\sum_{i\in{\Bbb
N}}{1\over 2^{i+1}}$ --- it follows that the construction below is 
not canonical. Obviously,
there are also less pathological differences between the different 
non-comparable discrete
representations \cite{C97}$^{ii}$.}

\smallskip
$\bullet$ $
\varphi_\alpha(p,\lambda):=
\bla
\ \ \alpha\ {\rm iff}\ P_p(\alpha)\geq{1\over
2^\lambda}+\sum_{i=1}^{\lambda-1}{\delta\left(\varphi_\alpha(p,i),\alpha\right)\over
2^i}\\
\neg\alpha\ {\rm otherwise}
\ela
$

\smallskip\noindent
One verifies that for $\mu(\lambda):={1\over 2^\lambda}$ we obtain the correct
probabilities in the resulting HMS-model.
This provides a discrete alternative for the above discussed ${\Bbb 
R}^3$-model for
spin-${1\over 2}$\,.  The model, including the projection $x_p$ 
remains the same although we
don't consider $[\alpha,\neg\alpha]$ as $\Lambda$ anymore.  Let 
$\lambda\in\Lambda':={\Bbb
N}$\,.  Set $x_n^\lambda:=\left(1-{n\over 2^\lambda}\right)\alpha+\left({n\over
2^\lambda}\right)\neg\alpha$ for $n\in{\Bbb Z}_{2^\lambda-1}$\,. For
$x_p\in[\alpha,x^\lambda_1[\cup[x^\lambda_2,x^\lambda_3[\cup[x^\lambda_4,x^\lambda_5[
\cup\ldots\cup[x^\lambda_{2^\lambda-1},\neg\alpha]$ we set
$\varphi'_\alpha(p,\lambda)=\alpha$\,, and
$\varphi'_\alpha(p,\lambda)=\neg\alpha$ otherwise. Then, for 
$\mu'_\circ:={\cal B}({\Bbb
N})\to[0,1]:\{\lambda\}\mapsto{1\over 2^\lambda}$ we obtain again quantum
probability\,.
Geometrically, this means that the values of
$\lambda\in\Lambda$\,, as compared to the first model where they 
represents points on the
diagonal, i.e., a continuous interval, or, again equivalently, 
decompositions of an interval
in two intervals, we now consider decompositions of an interval in
$2^\lambda$ equally long parts, of which there are only a discrete 
number of possibilities\,.
We refer to \cite{C97}$^{ii}$ for details and illustrations concerning.

\section{{\large UNITARY, ORTHO- AND PROJECTIVE STRUCTURE}}

In the above discussed ${\Bbb R}^3$ models, rotational symmetries 
where implicit in their
spatial geometry. However, in general the decompositions of 
measurements over $\mu:{\cal
B}(\Lambda)\to[0,1]$ go measurement by measurement so additional 
structure, if there is any,
has to be put in by hand.  It is probably fair to say that these 
contextual models only become
non-trivial and useful when encoding physical symmetries within the maps
$\varphi_\alpha$ in an appropriate manner\,.
For sake of the argument we will distinguish between three types of symmetries
that can be encoded, namely unitary, ortho- and projective ones.

\smallskip\noindent{\it i. Unitary symmetries\,:}
When considering quantum measurements with discrete
non-degenerated spectrum we can represent the outcomes $\{o_i\}_i$ by 
the corresponding
``eigenstates'' $\{p_i\}_i$ via spectral decomposition, i.e., there 
exists an injective map
${\cal B}(O_e)\rightarrow{\cal P}(\Sigma)$ for each $e\in{\cal E}$\,. 
Then, specification of
$\varphi:\Sigma\times\Lambda\to\{p_i\}_i$ and $\mu$ for one 
measurement $e_0\in{\cal E}$ fixes
it for any other
$e\in{\cal E}$ by symmetry: $\varphi_e=(U\circ\varphi\circ
U^{-1}):\Lambda\times\Sigma\to\{p_{e,i}\}_i$\,, where 
$U:\Sigma\to\Sigma$ is the unitary
transformation that satisfies $U(p_i)=p_{e,i}$\,, and $\mu_e=\mu$\,. 
This is exactly the
symmetry encoded in the above described ${\Bbb R}^3$-models\,.  Note 
in particular that in this
perspective the pairs
$(\alpha,\neg\alpha)$ and $\bigl(\neg\alpha,\neg(\neg\alpha)\bigr)$ 
should not be envisioned as
merely a change of names of the outcomes, but truly as putting the 
measurement device (or at
least its detecting part) upside down\,.\footnote{The attentive 
reader will note that it is at
this point that we escape the so-called hidden variable no-go 
theorems\,. They arise when
trying to impose contextual symmetries within the states of the 
system by requiring that
values of observables are independent of the chosen context, e.g., 
the proof of the
Kochen-Specker theorem. Our newly introduced variable 
$\lambda\in\Lambda$ follows contextual
manipulations in an obvious manner.} In this setting where we 
represent outcomes as states, the
assignment of an outcome can now be envisioned as a true change of state
$f_{e,\lambda}:\Sigma\to\Sigma\,(\supseteq
O_e):p\mapsto\varphi_e(p,\lambda)$\,,
as such allowing to describe the behavior of the system under concatenated
measurements.

\smallskip\noindent{\it ii. Projective symmetries\,:}
For non-degenerated quantum measurements, the outcomes
require representation by higher dimensional subspaces so 
identification in terms of
states now requires an injective map ${\cal B}(O_e)\rightarrow{\cal 
P}({\cal P}(\Sigma))$\,.
The behavior of states of the system under concatenated measurements 
then requires
specification of a family of ``projectors'' $\{\pi_T:\Sigma\to 
T\,|\,T\in O_e\}$\,, e.g.,
the orthogonal projectors $\pi_A:\Sigma\to A:p\mapsto A\wedge(p\vee 
A^\perp)$ on the
corresponding subspace
$A$ in quantum theory. The above discussed non-degenerated case fits 
also in this picture by
setting
$O_e\subseteq\bigl\{\{p\}\bigm|p\in\Sigma\bigr\}$ where now each 
$\pi_{\{p\}}:\Sigma\to\{p\}$
is uniquely determined (having a singleton codomain).

\smallskip\noindent{\it iii. Orthosymmetries\,:}
The existence of an orthocomplementation on the lattice of closed 
subspaces of a Hilbert space
provides a dichotomic representation for measurements which can be 
envisioned as a pair
consisting of a (to be verified) proposition $\alpha$ and its 
negation $\neg\alpha$\,, in
quantum theory yielding $\pi_{\neg A}:\Sigma\to A^\perp:p\mapsto 
A^\perp\wedge(p\vee
A)$\,. In terms of linear operator calculus we have $\pi_{\neg 
A}=1-\pi_A$\,, both of them
being orthogonal projectors.

\section{{\large REPRESENTING QUANTUM HISTORY THEORY}}

Although quantum history theory involves sequential measurements, one 
of its goals is to
remain an essentially dichotomic propositional theory.  This is 
formally encoded in a
rigid way in the ``History Projection Operator''-approach \cite{I94}. 
The key idea
here is that the form of logicality aimed at in \cite{I94}
represents faithfully in the Hilbert space tensor 
product.\footnote{At this point we mention
that in the study of sequential phenomena in the axiomatic quantum 
theory perspective on
quantum logic, sequentiality and compoundness both turn out to be 
specifications of a universal
causal duality \cite{CMS01}, as such providing a metaphysical 
perspective on the use of tensor
products both for the description of compound physical systems and 
sequential processes.}  Let
${\cal A}:=(\alpha_{t_i})_i$ be a (so-called homogeneous) quantum 
history proposition with
temporal support $(t_1, t_2,\ldots,t_n)$\,. Then, rather than 
representing this as a sequence
of subspaces  $(A_i)_i$ or projectors $(\pi_i)_i$ we will either 
represent ${\cal A}$ as a pure
tensor
$\otimes_iA_i$ in the lattice of closed subspaces of the tensor 
product of the corresponding
Hilbert spaces or as the orthogonal projector $\otimes_i\pi_i$ on 
this subspace.  The crucial
property of this representation is then that $\neg{\cal A}$ again 
encodes as a projector namely
$id-\otimes_i\pi_i$ \cite{I94}, clarifying the notations
$\pi_{\cal A}$ and $\pi_{\neg{\cal A}}$\,.  Moreover, if $\{{\cal A}^i\}_i$ is
a set of so-called disjoint history propositions, i.e., 
$\otimes_kA_k^i\perp\otimes_kA_k^j$ for
$i\not=j$\,, then, the history proposition that expresses the 
disjunction of $\{{\cal
A}^i\}_i$ sensu \cite{I94} is exactly encoded as the projector 
$\sum_i\otimes_k\pi_k^i$\,.
We get as such a kind of logical setting that is still encoded in 
terms of projectors.
Note that $\pi_{\neg{\cal A}}$ is not of the form $\otimes_i\pi_i$ 
but of the form
$\sum_i\otimes_k\pi_k^i$ breaking the structural symmetry between a 
proposition and its
negation in ordinary quantum theory\,.

We will now transcribe the observations in the two previous sections 
to this setting in order
to provide a contextual deterministic model for quantum history 
theory with discretely
originating probabilities.  One could say that we will apply a split 
picture in terms of
Schr\"odinger-Heisenberg, namely we assume that on the level of 
unitary evolution we apply the
Heisenberg picture such that we can fix notation without reference to 
this evolution, but for
changes of state due to measurement we will (obviously) express this 
in the state space.
When encoding outcomes in terms of states we need to consider $n$ copies
of $\Sigma$\,, encoding the trajectories due to the measurements. In 
view of the
considerations made above it will be no surprise that we will 
consider these trajectories as
of the form $\otimes_ip_i$ in the tensor product 
$\otimes_i\Sigma_i$\,. This will require the
introduction of the following ``pseudo-projector'':

\smallskip
$\bullet$ $
\pi_{\cal
A}^\otimes:\Sigma\to\otimes_i\Sigma_i:p\mapsto p^\otimes_{\cal
A}:=p\otimes\pi_1(p)\otimes\ldots\otimes
(\pi_{n-1}\circ\ldots\circ\pi_1)(p)
$\,.

\smallskip\noindent
Setting $\Sigma^\otimes_{\cal A}:=\pi^\otimes_{\cal 
A}[\Sigma]=\{p^\otimes_{\cal
A}|p\in\Sigma\}$ then $\pi_{\cal
A}^\otimes:\Sigma\to\Sigma^\otimes_{\cal A}$ encodes a bijective 
representation of $\Sigma$\,.
Noting that $P_p({\cal A}):=\langle\,p^\otimes_{\cal A}\,|\,\pi_{\cal 
A}\,p^\otimes_{\cal
A}\,\rangle$ is the probability given by quantum theory to obtain 
${\cal A}$\,, we then set
inductively for fixed
$\lambda\in{\Bbb N}$ that
$\varphi_{\cal A}(p,\lambda)=
{\cal A}$ if and only if:

\smallskip
$\bullet$ $\langle\,p^\otimes_{\cal A}\,|\,\pi_{\cal A}\,p^\otimes_{\cal
A}\,\rangle\geq{1\over 
2^\lambda}+\sum_{i=1}^{\lambda-1}{\delta\bigl(\varphi_{\cal
A}(p,i)\bigr)\over 2^i}$\ \

\smallskip\noindent
and
$\varphi_{\cal A}(p,\lambda)=\neg{\cal A}$ otherwise\,.
The outcome trajectories in case we obtain ${\cal A}$ are then given 
in terms of initial states
by
$(\pi_{\cal A}\circ\pi_{\cal A}^\otimes):\Sigma\to\otimes_iA_i$\,.
The value $\lambda\in{\Bbb N}$ can be envisioned
as follows. We assume it to be a number of contextual events, either 
real or virtual
depending on one's taste, and we assume that, given that some events 
already happened, the
chance of a next one happening is equal to the chance that it doesn't 
happen, so we actually
consider a finite number of probabilistically balanced consecutive 
binary decisive processes
where the result of the previous one determines whether we actually 
will perform the next
one\,. Unitary symmetries are induced in the obvious way as
tensored unitary operators $\otimes_iU_i$\,. This model then produces 
the statistical behavior
of quantum history theory.

The breaking of the structural symmetry between a proposition and its
negation manifestates itself in the most explicit way in the sense 
that when we have a
determined outcome $\neg{\cal A}$ we don't have a determined 
trajectory in our model ---
obviously one could build a fully deterministic model that also 
determines this by
concatenation of individual deterministic models (one for each 
element in the temporal
support), but we feel that this would not be in accordance with the 
propositional flavor a
history theory aims at. The negation
$\neg{\cal A}$ is indeed cognitive and not ontological with respect 
to the actual executed
physical procedure or, in other words, the system's context, and one 
cannot expect an
ontological model to encode this in terms of a formal duality.  Explicitly,
$\neg(A\otimes B)$ can be written both as
$({\cal H}\otimes\neg B)\oplus(\neg A\otimes B)$ and $(\neg A\otimes{\cal
H})\oplus(A\otimes\neg B)$ which clearly define different procedures 
with respect to imposed
change of state due to the measurement.
Even more explicitly, setting ${\cal
HPO}(\{{\cal H}_k\}_k):=\{\sum_i\otimes_kA_k^i|A_k^i\in{\cal L}({\cal H}_k),
\otimes_kA_k^i\perp\otimes_kA_k^j\
\mbox{\rm for}\ i\not=j\}$ for ${\cal L}({\cal H}_k)$ the lattice of 
closed subspaces of
${\cal H}_k$\,, the ``ontologically faithful hull'' of ${\cal
HPO}(\{{\cal H}_k\}_k)$ consists then of all ``ortho-ideals'' ${\cal 
OI}\bigl({\cal
HPO}(\{{\cal H}_k\}_k)\bigr):=$

\smallskip
$\bullet$ $\bigl\{\downarrow\![\{\otimes_kA_k^i\}_i]\bigm|A_k^i\in{\cal
L}({\cal H}_k),\otimes_kA_k^i\perp\otimes_kA_k^j\
\mbox{\rm for}\ i\not=j\bigr\}$

\smallskip\noindent
where $\downarrow\![-]$ assigns to a set of pure tensors all pure 
tensors in $\otimes_k{\cal
H}_k$ that are smaller than at least one in the given set, this with 
respect to the ordering in
${\cal L}(\otimes_k{\cal H}_k)$ --- the downset $\downarrow\![-]$ 
construction makes ${\cal
OI}\bigl({\cal HPO}(\{{\cal H}_k\}_k)\bigr)$ inherit the ${\cal 
L}(\otimes_k{\cal H}_k)$-order as
intersection. If a particular decomposition is
specified as an element of
${\cal OI}\bigl({\cal HPO}(\{{\cal H}_k\}_k)\bigr)$\,, what means 
full specification of the
physical procedure where summation over different sequences of pure 
tensors is now envisioned
as choice of procedure, we can provide a deterministic  contextual 
model, the choice of
procedure itself becoming an additional variable.
Conclusively, the HPO-setting ``looses'' part of the physical 
ontology that goes with an
operational perspective on quantum theory,\footnote{A choice that is 
motivated by the
traditional consistent history setting and its interpretation as well 
as by a particular
semantical perspective on quantum logic as a whole.} and as such, if 
we want to provide a
deterministic representation for general inhomogeneous history 
propositions sensu the one we
obtained for the homogeneous ones, we formally need to restore this 
part of the physical
ontology, e.g. as ${\cal OI}\bigl({\cal
HPO}(\{{\cal H}_k\}_k)\bigr)$\,.

\section{{\large FURTHER DISCUSSION}}

In this paper we didn't provide an answer and we even didn't pose a 
question.  We just
provided a new way to think about things, slightly confronting the 
usual consistency or
decoherence perspective for history theories. Even if one does not 
subscribe to the underlying
deterministic nature of the model it still exhibits what a minimal 
representation of the
indeterministic ingredients can be, as such representing it in a more 
tangible way. With
respect to the non-existence of a smallest representation, in view of 
other physical
considerations it could be that one of the constructible discrete models
presents itself as the truly canonical one, e.g.,  equilibrium or 
other thermodynamical
considerations, metastatistical ones, emerging from additional modelization.

\section*{{\large ACKNOWLEDGMENTS}}

We thank Chris Isham for useful discussions on the content of this paper.

\end{document}